\documentclass[12pt,a4paper]{article}
\usepackage{graphicx}
\usepackage{times}
\title{\bf Signatures of binary evolution processes in massive stars}
\author{Dany Vanbeveren\\
\vspace{1cm}\\
\normalsize  Astrophysical Institue, Vrije Universiteit Brussel, Pleinlaan 2, 1050 Brussels, Belgium\\ 
\normalsize dvbevere@vub.ac.be\\ 
\normalsize and \\
\normalsize GroupT Leuven Engineering College, \\
\normalsize Association KU Leuven, Vessaliusstraat 13, 3000 Leuven, Belgium}
\date{\mbox{}}
\begin{document}
\maketitle
\pagestyle{empty}
%
%
\def\bull{\vrule height .9ex width .8ex depth -.1ex}
\makeatletter
\def\ps@plain{\let\@mkboth\gobbletwo
\def\@oddhead{}\def\@oddfoot{\hfil\tiny\bull\quad
``The multi-wavelength view of hot, massive stars''; 39$^{\rm th}$ Li\`ege Int.\ Astroph.\ Coll., 12-16 July 2010 \quad\bull}%
\def\@evenhead{}\let\@evenfoot\@oddfoot}
\makeatother
%
%
\def\beginrefer{\section*{References}%
\begin{quotation}\mbox{}\par}
\def\refer#1\par{{\setlength{\parindent}{-\leftmargin}\indent#1\par}}
\def\endrefer{\end{quotation}}
%
%
{\noindent\small{\bf Abstract:} 
Before binary components interact, they evolve as single stars do. We therefore first critically discuss massive single star processes which affect their evolution, stellar wind mass loss and rotation in particular. Next we consider binary processes and focus on the effect of rotation on binary evolution and on the mass transfer during Roche lobe overflow. The third part highlights the importance of close pairs on the comprehension of the evolution of stellar populations in starburst regions.}
%
%
\section{Introduction}
Massive stars are among the most important objects in the Universe and many (most?) of them are formed in binaries.  A selection of observational and theoretical facts that illustrate the importance of binaries and the evolution of massive and very massive stars in clusters with special emphasis on massive binaries have been summarized in two recent review papers  (Vanbeveren, 2009 = paper I, and  2010).  The present written version of the Liege binary review  can be considered as an addendum of both papers. 

\section{The evolution of massive single stars}
Before one or two massive stars in a binary start to interact, they evolve as single stars do. We therefore first discuss briefly massive single star evolution.

\subsection{The effect of stellar wind mass loss}

The evolution of a massive star depends significantly on its mass loss by stellar wind. We may distinguish 4 stellar wind mass loss phases: the OB-type phase, the luminous blue variable (LBV) phase, the Wolf-Rayet (WR) phase and the red supergiant (RSG) phase. They were discussed in paper I but let me focus a bit more on the RSG-phase.

\subsubsection{The RSG stellar wind}

Single stars with an initial mass $\le$ 30-40 M$_\odot$ become RSGs and therefore RSG stellar wind dominates their further evolution. Most of the stellar evolutionary codes use the RSG wind formalism proposed by de Jager et al. (1988) however this formalism predicts the real RSG wind rates perhaps not better than a factor 5-10. These rates determine whether or not the massive star will lose most of its hydrogen rich layers, e.g. whether or not the massive star will become a WR star. This means that a large uncertainty factor in the RSG rates may seriously affect the theoretically predicted population of WR stars. It was shown in Vanbeveren (1991) that a 35 \% increase of the RSG rates (compared to the de Jager rates) is sufficient in order to obtain correspondence between the observed and theoretically predicted WR population in the Solar neighbourhood. This was worked out in more detail in Vanbeveren (1996) and in Vanbeveren et al. (2007). Note that also the Padua group (Salasnich et al., 1999) defended evolutionary computations of massive stars with larger RSG rates. 

Interestingly, Yoon \& Cantiello (2010) recently studied the evolution of massive stars with pulsation driven super-winds during the RSG phase. As far as the effect on massive star evolution is concerned, they essentially arrive at similar conclusions as in our work. 

Let me finally remark that all the massive star population studies performed in Brussels since 1996 account for these higher RSG rates (in particular the population of the different SN types, De Donder \& Vanbeveren (2003); the WR population, Vanbeveren et al., 1998a, Van Bever \& Vanbeveren, 2000, 2003; binaries and the chemical evolution of the Galaxy, De Donder \& Vanbeveren, 2004, etc.).  

\subsection{The effect of rotation}

The effects of rotation on the evolution of massive single stars has been studied very intensively in the past 2 decades by the Geneva group (see the contribution of George Meynet in the present proceedings).  One may distinguish three main effects: rotating stars have larger convective cores than non-rotating stars, rotational mixing is responsible for the transport of interior matter up to the surface, the faster the rotation the higher the stellar wind mass loss rate.

\subsubsection{Rotation, convective cores and rotational mixing}

The faster the initial rotation of a massive single star the larger the convective core, e.g. the effect of rotation on massive star cores is similar to convective core overshooting. Based on the observed rotation velocities of galactic O-type stars one concludes that they are born with an initial rotational velocity $\approx$ 200-300 km/s (however see also section 3.1). The calculations of the Geneva team then let us conclude that on average moderate convective core overshooting mimics the average effect of rotation on convective cores.

Due to rotational mixing, matter from the interior may reach the surface layers. This mixing process only slightly modifies the overall evolution of the massive star, but it may alter the chemical abundances of the surface layers and a comparison with observed abundances may decide upon the effectiveness of rotational mixing. Unfortunately the correspondence between theoretical prediction and observations is rather poor (Hunter et al.,  2008). Binaries may help (Langer et al., 2008) but magnetic fields may be needed as well (see the contribution of I. Brott in the present proceedings). A general warning may be appropriate: comparing the Hunter diagram with theoretical prediction involves population synthesis. In Brussels we have been studying massive star and binary populations for about two decades and we frequently found poor correspondence with observations, but in many cases, after second thoughts we concluded that observational bias was one of the main reason. Selecting well observed stars/binaries and trying to explain them may be at least as useful as overall population synthesis. OBN binaries are known quite some time and I like to call your attention to the interesting system HD 163181. It is a BN0.5Ia  + OBN binary (Hutchings, 1975; Josephs et al., 2001) with a period P = 12 days and masses 13 M$_\odot$ + 22 M$_\odot$. The 13 M$_\odot$ primary has the luminosity of a 30 M$_\odot$ main sequence star and is at least 1.5 mag brighter than the 22 M$_\odot$ secondary. Binary evolution then reveals that the primary must be a core helium burning star at the end of RLOF (Vanbeveren et al., 1998b). The nitrogen enhancements are most probably due to binary mass exchange and more observations of this system may prove to be very instructive. Interestingly, the 13 M$_\odot$ mass loser has all the properties of WR stars in binaries (except may be the Teff) but it is not a WR star. We suspect that in the very near future the supergiant will turn into a WR star. 
 
An embarrassing problem of the present rotating massive star evolutionary models is that they produce pulsars that are spinning too fast by at least an order of magnitude (Heger et al., 2000) and also here the coupling between rotation and magnetic fields may be the solution. 

Some massive stars are known to rotate very fast, close to break up (the Be stars but also some O-type stars). Due to the combined action of convective core growth and rotational mixing, stars that rotate close to the critical velocity are expected to evolve quasi-homogeneously (Maeder, 1987) and their evolution is quite different from the 'normal' evolution of massive stars.  An important question is how these stars became extremely rapid rotators. Many rapid rotators are known to be binary components or former binary components and therefore we will come back to this in section 3.1.

\subsubsection{Rotation and the stellar wind mass loss rate}

One of the main evolutionary effects of rotation is related to the effect of rotation on the stellar mass loss. A first attempt to link rotation and mass loss was proposed by Langer (1997) but Glatzel (1998) showed that the proposed relation may be not correct due to the fact that it does not account for the effect of gravity darkening (von Zeipel, 1924). An alternative and attractive formalism has been derived by Maeder and Meynet (2000) where the effect of gravity darkening was taken into account. This relation demonstrate that for most of the massive stars (with an average initial rotational velocity of $\approx$ 200-300 km/s) the increase of the stellar wind mass loss with respect to the non-rotating case is very modest. The increase is significant for stars with a large Eddington factor $\Gamma$ (e.g., stars with an initial mass $\ge$ 40 M$_\odot$) that are rotating close to critical. However, the following remarks are appropriate: there are no observations yet to sustain the relation proposed by Maeder and Meynet (Puls et al., 2010). Even more, one may wonder whether or not rotation can be a significant mass loss driver, since even at critical rotation, the rotational energy is at most half the escape energy of a massive star (Owocki, 2010, private communication). 

\section{The evolution of massive binaries}

The main differences between single star evolution and the evolution of the same star when it is a binary component are related to  the Roche lobe overflow (RLOF) process and to binary processes which determine the rotation rate of the star.  

\subsection{Rotation and binaries}

Due to tidal interaction the massive primaries of most of the binaries are expected to be slow rotators. Only in very short period systems (P = 1-2 days) it can be expected that massive primaries are rapid rotators for which the evolution proceeds quasi-homogeneously (De Mink, 2010). 

In binaries where the RLOF of the primary is accompanied by mass transfer towards and mass accretion onto the secondary, the secondary spins-up and very rapidly rotational velocities are reached close to the critical one (Packet, 1981). This happens in  systems where the RLOF occurs when the outer layers of the primary are mainly in radiative equilibrium (Case A / Case Br systems). Population synthesis predicts that many Be stars are formed this way (Pols and Marinus, 1994; Van Bever and Vanbeveren, 1998). The latter two studies illustrate that one may expect many Be stars with a subdwarf (sdO) or white dwarf (WD) companion. The high temperature of these companions makes them very hard to detect and this may be the reason why so few are known at present ($\phi$ Per is an exception, see section 3.3.1). The Be-components in Be-X-ray binaries form an interesting subclass of the Be sample because here we have all reasons to believe that binary action has been important in the formation of the Be star. Many Be single stars are also expected to form via binary mass transfer. The reason is that the supernova explosion of a massive primary disrupts the binary in most of the cases. This means that the fact that many Be stars have a neutron star companion means that even more Be single stars have had a similar past as the Be stars in the Be-X-ray binaries.

The optical components of the standard high mass X-ray binaries are former binary secondaries where mass and angular momentum accretion may have occured. The mass and helium discrepancy for single stars discussed by Herrero et al. (1992) is also visible in the optical component of the X-ray binary Vela X-1 (Vanbeveren et al., 1993) and we proposed  {\it the accretion induced full mixing model}.  The idea was the following: due to mass and angular momentum accretion, mass gainers spin-up. This may induce efficient mixing. We simulated this possibility with our evolutionary code by fully mixing the mass gainer and, after the mass transfer phase, following the further evolution of the mixed star in a normal way. In this way we were able to explain the helium and mass discrepancy in Vela X-1. The more sophisticated mass gainer models of Cantiello et al. (2007) demonstrate that our simplified models are not too bad.

\begin{figure}[h]
\centering
\includegraphics[width=10cm]{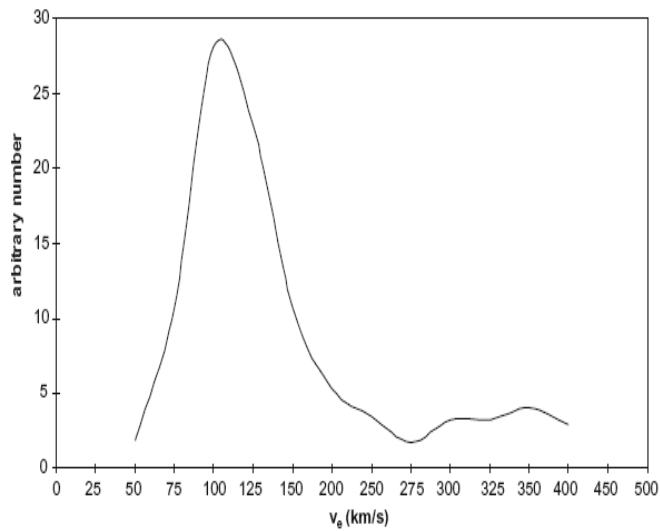}
\caption{The probable v$_{rot}$ distribution of O-type stars using the data of Penny (1996) (from Vanbeveren et al., 1998).\label{fig_1}}
\end{figure}

Starting from the observed $v_{rot}sin$i data of O-type stars of Penny (1996) Figure 1 shows the probable distribution of rotational velocities $v_{rot}$ of O-type stars (from Vanbeveren et al., 1998a) and illustrates that many O-type stars are relatively slow rotators, corresponding to an initial average rotational velocity for O-type stars of $\le$ 200 km/s for which indeed the effect of rotation on their evolution is rather modest. The figure also shows that there is a subset of very rapid rotators. However, many of these rapid rotators are runaway stars (they have a space velocity $\ge$ 30 km/s) and this may indicate that these stars were former binary components, e.g. they became rapid rotators due to the mass transfer process in a binary and they became runaways due to the supernova explosion of their companion, or they became rapidly rotating runaways due to stellar dynamics in dense stellar clusters, in which case they were former binary members as well but their formation was governed by star merging. An interesting test bed for this type of process may be $\zeta$ Pup which is indeed a rapidly rotating runaway. Note that Mokiem et al. (2006) obtained rotational velocities of 21 OB dwarfs in the SMC and concluded that the average  $v_{rot}$=160-190 km/s. Since massive dwarfs are stars close to the zero age main sequence, this average value is indicative for the average initial rotation velocity of OB-type stars. Remark that this value is not significantly different than the initial average value of Galactic O-type stars whereas, similar as in the Galactic sample, the most rapid rotators in the SMC are runaway stars. 

All in all, it looks to me that rotation is important for massive star evolution but perhaps mainly within the framework of binaries in combination with stellar dynamics in young dense clusters.  

\subsection{The Roche lobe overflow process}

When the RLOF starts when the mass loser has a convective envelope (Case Bc and Case C), the mass loss process happens on the dynamical timescale and a common envelope forms. It can be expected that the common envelope is lost as a superwind where most  of the energy is supplied by orbital decay and it stops when the two components merge or, when merging of the two stars can be avoided, when most (but not all) of the hydrogen rich layers of the mass loser are removed. This phase is so rapid that it is unlikely that  mass accretion plays an important role for the evolution of the secondary star and therefore the latter may not become a rapid rotator.  

When the RLOF starts when the mass loser has a radiative envelope (Case A and Case Br), the mass loss process happens on the Kelvin-Helmhotz time scale of the loser and when the initial mass of the gainer is not too much smaller than the initial mass of the loser, mass transfer and mass accretion becomes possible. 

The evolution of the mass loser in Case Br and in most of the Case A massive binaries is very straightforward: due to RLOF the redward evolution of the loser is avoided (e.g., massive primaries in Case A or Case Br binaries do not become red supergiants). The RLOF stops when the loser has lost most (but not all) of its hydrogen rich layers and helium starts burning in the core. At that moment the loser resembles a WR-like star (when the mass is large enough the WR-like star is expected to be a guinine WR star with hydrogen in its atmosphere, typically X = 0.2-0.3). 

The evolution of the mass gainer in Case A and Case Br binaries is governed by mass and angular momentum accretion and rotation plays a very important role (see the previous subsection).

An important question is whether or not the RLOF in Case A or Case Br binaries is quasi-conservative. Let me first remark that removing matter out of a binary at a rate which is similar to the rate at which the primary loses mass, requires a lot of energy, much more than the intrinsic radiation energy of the two components which is in most cases only sufficient in order to drive a modest stellar wind. The Utrecht group promoted a massive binary model where extensive mass loss from the system happens when the mass gainer has been spun-up by mass and angular momentum transfer and reaches a rotational velocity close to the critical one (Petrovic et al., 2005; De Mink, 2010). However, as discussed already in subsection 2.2.2, rotation is not an efficient mass loss driver whereas even at break up, the rotational energy is at least a factor 2 too small compared to the required escape energy. Van Rensbergen et al. (2008) proposed the following model: the gasstream during RLOF forms a hot spot either on the surface of the star when the gasstream hits the mass gainer directly, or on the Keplerian disc when mass transfer proceeds via a disc. The radiation energy of the hot spot in combination with the rotational energy of the spun-up mass gainer can then drive mass out of the binary. The following illustrates that this model is probably not a process that removes from the binary a significant amount of mass  lost by the loser. For the sake of simplicity, let us neglect rotation because as was stated already before, rotation is not an efficient mass loss driver whereas even at critical break up, the rotation energy is too small compared to the escape energy of a massive star. The radiation energy $L_{acc}$ generated by the accretion of the gasstream is given by

\begin{equation}
L_{acc} = G\frac{\dot{M}_{acc}\,M}{R}
\end{equation}

\noindent where we neglect the fact that the gasstream does not originate at infinity but at the first Lagrangian point (it can readily be checked that this assumption does not significantly alter our main conclusions). $\dot{M}_{acc}$ is the mass accretion rate, M and R are the mass and the radius of the gainer. When $\eta$ is the fraction of $L_{acc}$ that is effectively transformed into escape energy, it follows that

\begin{equation}
\eta\ L_{acc} = \frac{1}{2}\dot{M}_{out} v^{2}_{esc}
\end{equation}

\noindent with $\dot{M}_{out}$ the binary mass loss rate and $v_{esc}$ the binary escape velocity. In order to get a first order estimate of the binary mass loss rate, we can replace $v_{esc}$ by the escape velocity of the mass loser only. The foregoing equations then result into

\begin{equation}
\dot{M}_{out}=\eta\ \dot{M}_{acc}
\end{equation}

\noindent Detailed hydrodynamic calculations of stellar winds reveal that the efficiency factor for converting radiation energy into kinetic energy is of the order of 0.01 and 0.001 (Nugis \& Lamers, 2002). So, unless the efficiency is much higher, equation (3) then illustrates that in general the accretion energy may cause mass loss out of a binary but this loss is much smaller than the mass accretion rate. Of course when the gainer rotates at the critical velocity, mass accretion on the equator will not happen. Possibly matter will pile up arround the gainer, accretion may happen on the rest of the star, or matter may leave the binary through the second Lagrangian point L2 as illustrated in Figure 2. When this happens the variation of the orbital period can be calculated in a straightforward way (e.g., Vanbeveren et al., 1998b). Note that in most of the population studies performed by different groups the effect of a non-conservative RLOF is investigated using this L2 model. Let me finally remark that mass that leaves the binary through the decretion/accretion disk of the gainer when the gainer rotates at the break up velocity, takes with the specific orbital angular momentum of the gainer but also the specific rotational angular momentum of the equator of the gainer. Interestingly, the sum of both momenta is roughly equal to the specific angular momentum of the L2 point.

\begin{figure}[h]
\centering
\includegraphics[width=10cm]{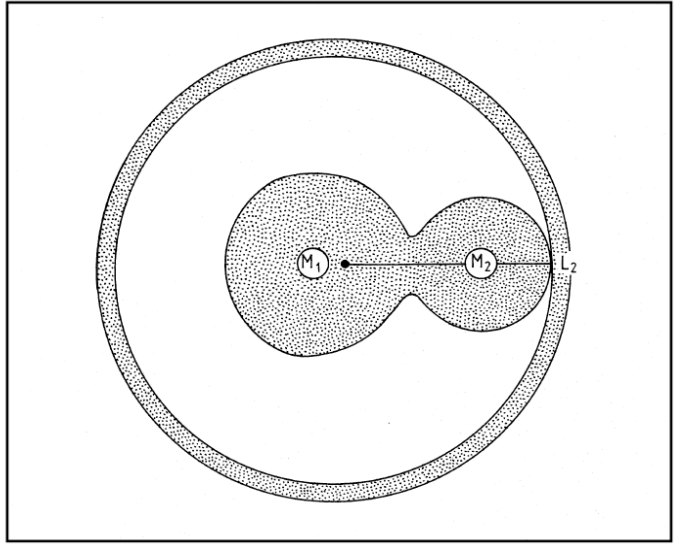}
\caption{Mass loss from the system through L2 and the formation of a ring around the binary, during the RLOF of the primary in a Case A/ Case Br binary.\label{fig_2}}
\end{figure}

\subsection{Some interesting observed test beds of the Roche lobe overflow process}

As discussed in the previous subsection, from theoretical considerations it is unclear whether or not Case A / Case Br evolution in massive binaries is quasi-conservative or not. Are there observed binaries that can be considered as RLOF test beds and allow us to say something about the RLOF? The best candidates are post-RLOF binaries or binaries at the end of RLOF where one can try to fit evolutionary models of binaries where the evolution of both components is followed simultaneously adopting different efficiency values for the mass transfer process. We did this for a number of Galactic massive binaries (Vanbeveren et al., 1998b) and de Mink et al. (2007) for SMC binaries. Here we reconsider two most interesting systems.

\subsubsection{$\phi$ Per}

$\phi$ Per is a sdO6 + B0.5Ve binary with a period = 126 days. It is a post-RLOF binary where the subdwarf O6 star has been the mass loser (thus evolutionary speaking this is the primary although it is by far the less luminous component = visual luminosity) and the Be star is the mass gainer (the secondary). The fact that the secondary is a rapidly rotating Be star is indicative that mass transfer has played an important role. There are two different studies where the masses of the components were determined, e.g., 1.7 M$_\odot$ +  17.3 M$_\odot$ (Bozic et al., 1995) and 1.14 M$_\odot$ + 9.3 M$_\odot$ (Gies et al., 1996). Accounting for the period of the binary, both sets of masses imply that the previous RLOF was quasi-conservative (Vanbeveren et al., 1998b).

\subsubsection{RY Scuti}

The massive binary RY Scuti may be a key system for the discussion whether or not the RLOF in massive binaries is conservative. The spectroscopic study of Grundstrom et al. (2007) reveals that it is a O9.7Ibpe + B0.5I binary with a period = 11.2 days and masses 7 M$_\odot$ + 30 M$_\odot$. The O-type supergiant is the most luminous component and comparison with evolutionary models of binaries reveals that it is most probably a core helium burning star near the end of RLOF with a significantly reduced surface hydrogen content (X = 0.3-0.4). Similarly as the supergiant in HD 163181 (section 2.2.1)we predict that the O-type supergiant will soon become a WR star. If the masses are correct then this system is an illustration of a massive binary where the RLOF was quasi-conservative for the following reasons. From evolution it follows that the 7 M$_\odot$ star comes from a star with initial mass $\le$ 20 M$_\odot$ that lost $\le$ 13 M$_\odot$ by RLOF. Since the initial mass of the secondary (mass gainer) must have been $\le$ 20 M$_\odot$ as well obviously, it must have accreted at least 10 M$_\odot$ of the $\le$ 13 M$_\odot$ lost by the loser in order to become a 30 M$_\odot$ star, e.g. the mass accretion efficiency must have been at least 80\% and we call this quasi-conservative. Note that the observations of Grundstrom et al.(2007) seem to indicate that there is some circum-binary material, but it is clear that the model discussed above does not contradict this fact.    

\subsubsection{The stellar wind and binaries}

Spherically symmetric stellar wind mass loss of one or both components in a binary increases the binary period and decreases the amount of mass lost by RLOF when the stellar wind mass loss happens before the latter. Stars with an initial mass $\ge$ 30-40 M$_\odot$ are losing their hydrogen rich layers by an O-type wind and by LBV-type mass loss. When  a star like that is a binary component with a period that is large enough so that the RLOF would start after these O and LBV type mass loss processes, the RLOF will not happen (the 'LBV scenario' of massive close binaries as it was introduced in Vanbeveren, 1991).

Our evolutionary computations of massive single stars but with our preferred RSG wind rates (section 2.1.2) predict that stars in the mass range 15-20  M$_\odot$ up to 30-40 M$_\odot$ lose sufficient hydrogen rich layers so that in the HR-diagram they turn back and become WR or WR-like stars. It is clear that when such a star is a binary member with a period that is large enough so that this RSG mass loss process starts before the Roche lobe overflow (RLOF) starts, the RLOF will be avoided. This means that what is called Case C evolution of binaries does not happen when the primary star has an initial mass $\ge$ 15-20 M$_\odot$ (the RSG scenario of massive binaries, Vanbeveren, 1996).

The evolutionary computations with  our preferred RSG rates also predict that stars in the mass range 10 M$_\odot$ and 15-20 M$_\odot$ lose a few M$_\odot$ during the RSG phase. When such a star is the primary of a binary with a period such that it will evolve through a common envelope (CE) phase, the total mass lost due to the latter CE process may be reduced due to the RSG stellar wind mass loss. The binary $\upsilon$ Sgr (Upsilon Sagittarii) may have evolved this way and may be an interesting test bed for the effect of RSG mass loss on binary evolution. It consists of an  A-type supergiant + B4-7 main sequence star. The period = 138 days and the present masses are  2.5 M$_\odot$ +  4 M$_\odot$. The A supergiant is extremely hydrogen deficient, core helium/helium shell burning, with Log L/L$_{\odot}$ = 4.6. Evolution predicts that the initial masses were 12 M$_\odot$ + 4 M$_\odot$. The binary is post-common envelope but to understand its evolution (in particular the period evolution) one has to accept that RSG stellar wind mass loss has been important and lowered the importance of common envelope mass loss (see also  Vanbeveren et al., 1998b).

\section{Close pairs - key to comprehension of the evolution of stellar populations in starburst regions}

Mass transfer and mass accretion during a canonical RLOF in Case A/Br binaries  is responsible for the formation of a blue straggler sequence in young clusters (Pols and Marinus, 1994; Van Bever and Vanbeveren, 1998, see Figure 3). Since these blue stragglers are mass gainers or binary mergers, it can be expected that they are rapid rotators, e.g., a cluster of slow rotators but with a significant population of close binaries will become populated with rapid rotators.     

\begin{figure}[h]
\centering
\includegraphics[width=10cm]{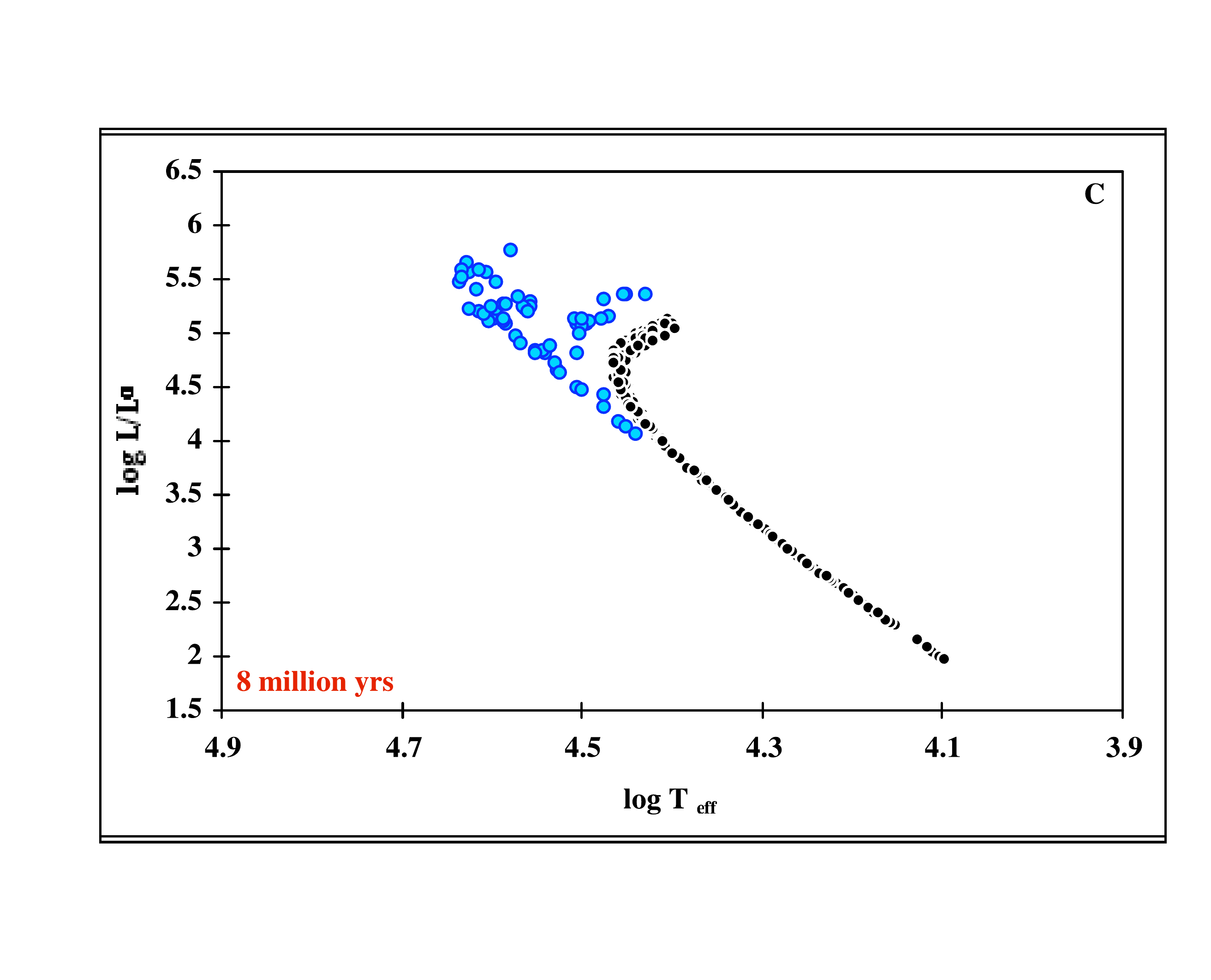}
\caption{A typical starburst with primordial binaries after 8 Myr; the blue stars are rapidly-rotating mass gainers or binary mergers (from Van Bever and Vanbeveren, 1998).\label{fig_3}}
\end{figure}

Starburst99 is an interesting spectral synthesis tool to estimate all kinds of properties of starburst regions where only integrated spectra are available. However, it should be noted that this tool only accounts for the properties of single stars. The effects of binaries on the spectral synthesis of starbursts has been studied in Brussels: the effects of binaries on the evolution of W(H$_\beta$) (Van Bever \& Vanbeveren, 1999), the effect of binaries on the evolution of UV spectral features in massive starbursts (Belkus et al., 2003), the effect of binaries on WR spectral features in massive starbursts (Van Bever \& Vanbeveren, 2003), hard X-rays emitted by starbursts with binaries (Van Bever \& Vanbeveren, 2000). Note that Brinchmann et al. (2008) investigated the spectral properties of WR galaxies in the Sloan Digital Sky Survey and concluded that a comparison with theoretical population synthesis leeds to the conclusion that binaries are necessary.

\subsection{Intermezzo 1}

Vanbeveren (1982) discussed a possible relation between the maximum stellar mass in a cluster and the total cluster mass. It was concluded that {\it the integrated galaxial stellar IMF should be steeper than the stellar IMF}. This has been worked out in more detail about 20 years later by Kroupa and Weidner (2003) and Weidner and Kroupa (2006) who essentially arrived at the same conclusion.

A consequence of the fact that the mass of the most massive star in a cluster correlates with the cluster mass, is that it is possible that also the mass ratio distribution of the most massive binary population in the cluster correlates with the cluster mass. To illustrate, suppose that the cluster mass indicates that the maximum stellar mass/total binary mass is 50 M$_\odot$, then one may expect binaries like 40 M$_\odot$ + 10 M$_\odot$ or 30 M$_\odot$ + 20 M$_\odot$ etc., but not 40 M$_\odot$ + 30 M$_\odot$.   
 
\subsection{Intermezzo 2}

Could it be that stars form in isolation? The origin of massive O-type field stars has been studied by De Wit et al., 2004. The authors proposed the following procedure in order to find a candidate O-type star that may have formed in isolation : take a non-runaway O-type field star and look for young clusters within 65 pc from the O-star. The value 65 pc was obtained by assuming that the lifetime of an O-type star is $\le$ $10^7$ yr. This is true for single stars however, the lifetime of an O-type star in a binary may be 2-3 times larger and therefore it may be necessary to look for young clusters within 200 pc. The model goes as follows: a 12 M$_\odot$ + 9 M$_\odot$ binary is dynamically ejected from a dense cluster with a velocity of 6 km/s. After 30 million years, when the binary is 200 pc away from its parent cluster, the 12 M$_\odot$ primary starts its RLOF. A quasi-conservative RLOF turns the 9 M$_\odot$ secondary into a 19 M$_\odot$ rejuvenated O-type star. When the primary remnant finally explodes as a supernova, the 19 M$_\odot$ O-type star most likely becomes a single star but the magnitude and direction of its space velocity may have changed completely, even so that its direction does not hit the parent cluster any longer. 

\subsection{Stellar dynamics in young dense star clusters}

Ultra Luminous X-ray sources (ULX) are point sources with X-ray luminosities up to 10$^{42}$ erg s$^{-1}$. MGG-11 is a young dense star cluster with Solar type metallicity $\sim$200 pc from the centre of the starburst galaxy M82, the parameters of which have been studied by McCrady et al. (2003). A ULX is associated with the cluster. When the X-rays are due to Eddington limited mass accretion onto a black hole (BH) it is straightforward to show that the mass of the BH has to be at least 1000 M$_\odot$. However how to form a star with Solar metallicity and with a mass larger than 1000 M$_\odot$? Mass segregation in a dense young cluster associated with core collapse and the formation of a runaway stellar collision process was promoted by Portegies Zwart et al. (2004). Note that the latter paper mainly addressed the dynamical evolution of a dense cluster but the evolution of the very massive stellar collision product was poorly described. 

\begin{figure}[h]
\centering
\includegraphics[width=10cm]{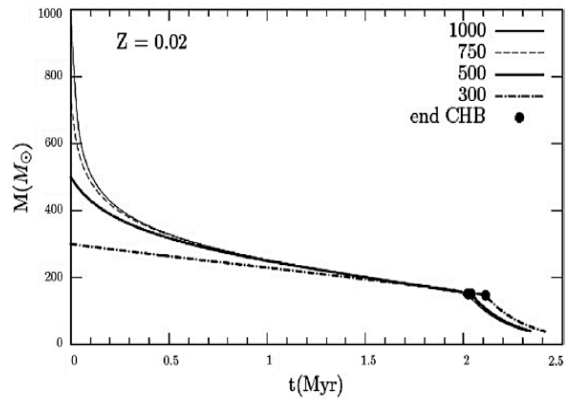}
\caption{The mass evolution during core-hydrogen-burning and during core-helium-burning of very massive stars (Solar metallicity) with an initial mass 300 M$_\odot$, 500 M$_\odot$, 750 M$_\odot$ and 1000 M$_\odot$ (from Belkus et al., 2007).\label{fig_3b}}
\end{figure}

\begin{figure}[h]
\centering
\includegraphics[width=10cm]{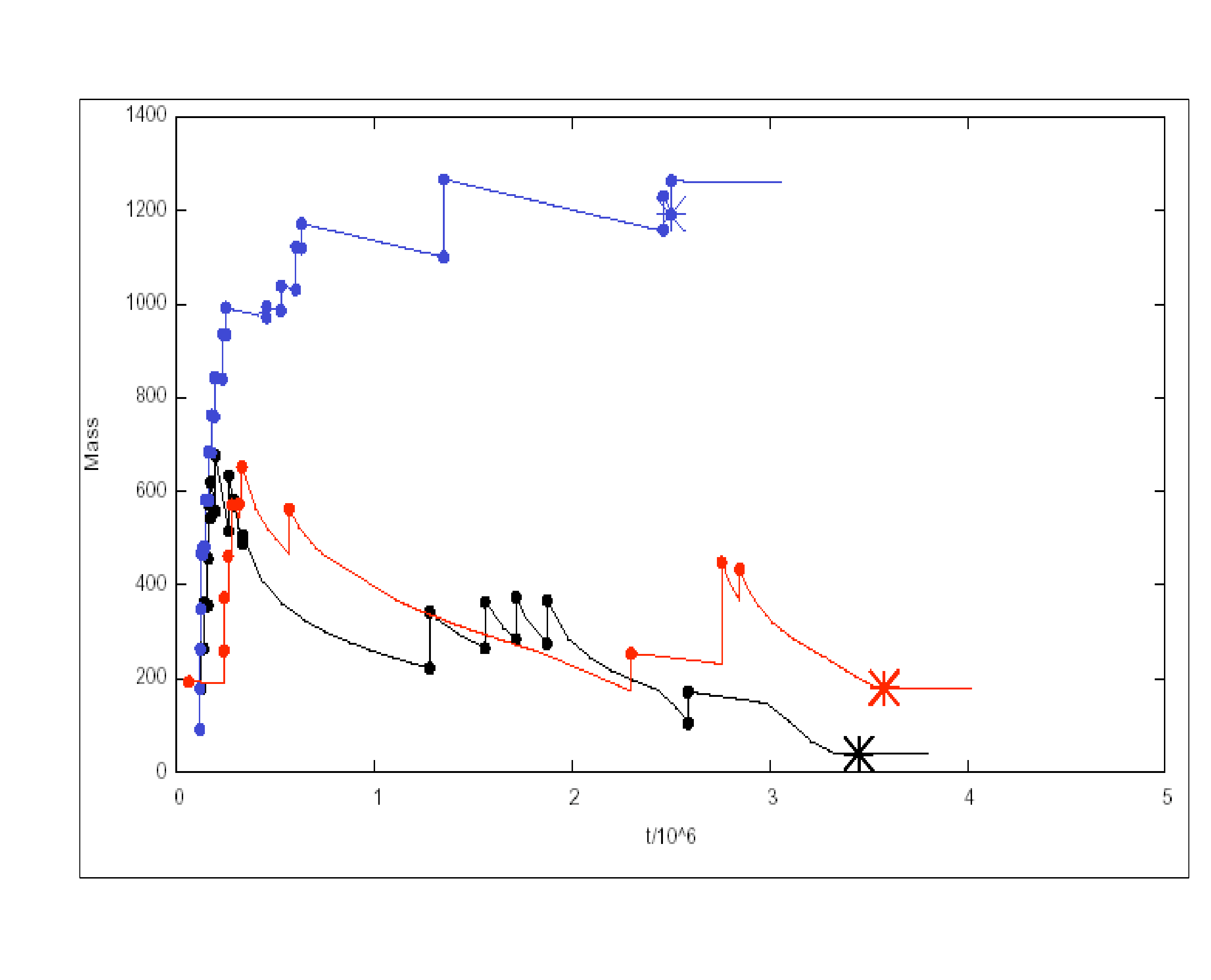}
\caption{The mass evolution of the collision runaway object in a MGG-11 type cluster. The curve with the largest final mass (Blue curve) = simulation with small mass loss, similar to the results of Portegies Zwart and McMillan (2002), the curve with the lowest final mass (Black curve) = simulation with the stellar wind mass loss formalism as discussed in Belkus et al. (2007) for Solar type metallicity, the curve with the second largest final mass (Red curve) = same as the black curve but for a SMC type metallicity (from Vanbeveren et al., 2009).\label{fig_4}}
\end{figure}

The evolution of very massive stars has been studied in detail by Belkus et al. (2007) and Yungelson et al. (2008) and it was concluded that stellar wind mass losses during core hydrogen burning and core helium burning are very important (Figure 4). Belkus et al. presented a convenient evolutionary recipe for such very massive stars, which can easily be implemented in an N-body dynamical code. Our N-body code which includes this recipe has been described in Belkus (2008) and in Vanbeveren et al. (2009) and applied in order to simulate the evolution of MGG-11. In Figure 5 we show the evolution of the runaway stellar collision object for MGG-11 predicted by our code. The blue simulation is performed assuming a similar stellar wind mass loss formalism for very massive stars as the one used by Portegies Zwart et al. (2004). It can readily be checked that our simulation is very similar as the one of the latter paper and this gives some confidence that our N-body-routine is working properly. We then repeated the N-body run but with our preferred evolutionary scheme for very massive stars (the black run in Figure 5).  Our main conclusion was  the following:

\medskip
{\it \noindent Stellar wind mass loss of massive and very massive stars does not prevent the occurrence of a runaway collision event and the formation of a very massive star in a cluster like MGG-11, but after this event stellar wind mass loss during the remaining core hydrogen burning phase is large enough in order to reduce the mass again and the formation of a BH with a mass larger than $\sim$75 M$_\odot$ is rather unlikely.} 

\medskip

Similar conclusions were reached by Glebbeek et al. (2009) for MGG-11 and by Chatterjee et al. (2010) for the Arches cluster, although in both studies cluster dynamics and the evolution of very massive stars are not linked self-consistently.

Our simulations then promote the model  for the ULX in MGG-11 where the X-rays are due to super-Eddington accretion onto a stellar mass BH, a model that seems to be a most probable model for many of these  systems (Gladstone et al., 2009).

We also made a simulation for a MGG-11 like cluster but where the metallicity is significantly smaller. As can be noticed from Figure 5 (the red simulation) the formation of a BH with a mass of a few 100 M$_\odot$ is possible and this is of course a direct consequence of our adopted dependence of the stellar wind mass loss rate on the metallicity. From this simulation we are inclined to conclude that, if the progenitors of globular clusters were massive starbursts in the beginning, it is not unlikely that an intermediate mass black hole formed as a consequence of mass segregation and core collapse in a dense massive cluster.

%
%
\footnotesize
\beginrefer

\refer Belkus, H., 2008, PhD Thesis (Vrije Universiteit Brussel).

\refer Belkus, H., Van Bever, J., Vanbeveren, D., 2007, ApJ 659, 1576.

\refer Belkus, H., Van Bever, J., Vanbeveren, D., van Rensbergen, W., 2003, A\&A 400, 429.

\refer Bozic, H., Harmanec, P., Horn, J., Koubsky, P. et al., 1995, A\&A 304, 235.

\refer Brinchmann, J., Kunth, D., Durret, F., 2008, A\&A 485, 657.

\refer Cantiello, M., Yoon, S.-C., Langer, N., Livio, M., 2007, A\&A 465, 29.

\refer Chatterjee, S., Goswami, S., Umbreit, S., Glebbeek, E., et al., 2009, arXiv0911.1483C

\refer De Donder, E., Vanbeveren, D., 2003, New Astronomy 8, 817.

\refer De Donder, E., Vanbeveren, D., 2004, New Astronomy Reviews 48, 861.

\refer de Jager, C., Nieuwenhuijzen, H., van der Hucht, K.A., 1988, A\&ASuppl 72, 259.

\refer de Mink, S. E., Pols, O. R., Hilditch, R. W., 2007, A\&A 467, 1181.

\refer De Mink, S.E., 	PhD Thesis (Utrecht University). 

\refer de Wit, W. J., Testi, L., Palla, F., Vanzi, L., Zinnecker, H., 2004, A\&A 425, 937.

\refer Gies, D.R., Thaller, M.L., Bagnuolo, W.G. Jr., Kaye, A.B., et al., 1996, Bull. American Astron. Soc. 28, 1373.

\refer Gladstone, J., Roberts, T.P., Done, C., 2009, MNRAS 397, 1836.

\refer Glatzel, W., 1998, A\&A 339, L5.

\refer Glebbeek, E., Gaburov, E., de Mink, S. E., Pols, O. R., Portegies Zwart, S. F., 2009, A\&A 497, 255.

\refer Grundstrom, E. D., Gies, D. R., Hillwig, T. C., McSwain, M. V., et al., 2007, ApJ 667, 505.

\refer Heger, A., Langer, N., Woosley, S. E, 2000, ApJ 528, 368.

\refer Herrero, A., Kudritzki, R. P., Vilchez, J. M., Kunze, D., et al., 1992, A\&A 261, 209.

\refer Hunter, I., Lennon, D.J., Dufton, P.L., Trundle, C., et al., 2008, A\&A 479, 541.

\refer Hutchings, J.B., 1975, ApJ 200, 122.

\refer Josephs, T. S., Gies, D. R., Bagnuolo, W. G., Jr.; Shure, M. A., et al., 2001, PASP 113, 957.

\refer Kroupa, P., Weidner, C., 2003, ApJ 598, 1076.

\refer Langer, N., 1997, in 'Luminous Blue Variables: Massive Stars in Transition', eds. Nota, A., Lamers, H., ASP Conf. Series 120, p. 83.

\refer Langer, N., Cantiello, M., Yoon, S.-C., Hunter, I., et al., 2008, IAUS 250, 167.

\refer Maeder, A., 1987, A\&A 178, 159.

\refer Maeder, A., Meynet, G., 2000, A\&A 361, 159.

\refer McCrady, N., Gilbert, A.M., Graham, J.R., 2003, ApJ 596, 240.

\refer Mokiem, M. R., de Koter, A., Evans, C. J., Puls, J., et al., 2006, A\&A 456, 1131.

\refer Nugis, T., Lamers, H., 2002, A\&A 389, 162.

\refer Packet, W., 1981, A\&A 102, 17.

\refer Penny, L.R., 1996, PhD Thesis (Georgia State University).

\refer Petrovic, J., Langer, N., van der Hucht, K. A., 2005, A\&A 435, 1013.

\refer Pols, O.R., Marinus, M., 1994, A\&A 288, 475.

\refer Portegies Zwart, S. F., Baumgardt, H., Hut, P., Makino, J. McMillan, S., L. W., 2004, Nature 428, 724.

\refer Puls, J., Sundqvist, J. O., Rivero Gonz‡lez, J. G., 2010, arXiv1009.0364P

\refer Salasnich, B., Bressan, A., Chiosi, C., 1999, A\&A 342, 131.

\refer Van Bever, J., Belkus, H., Vanbeveren, D., Van Rensbergen, W., 1999, New Astron. 4, 173.

\refer Van Bever, J., Vanbeveren, D., 1998, A\&A 334, 21.

\refer Van Bever, J., Vanbeveren, D., 2000, A\&A 358, 462.

\refer Van Bever, J., Vanbeveren, D., 2003, A\&A 400, 63.

\refer Van Rensbergen, W., De Greve, J.P., De Loore, C., Mennekens, N., 2008, A\&A 487, 1129.

\refer Vanbeveren, D., 1982, A\&A 115, 65.

\refer Vanbeveren, D., 1991, A\&A 252, 159.

\refer Vanbeveren, D., 1996, in 'Evolutionary processes in binary stars'', eds. Wijers, R.A.M.J. and Davies, M.B., ASI Series, vol. 477, Dordrecht: Kluwer Academic Publishers, p. 155

\refer Vanbeveren, D., 2009, New Astron. Reviews 53, 27.

\refer Vanbeveren, D., 2010,  in 'Star clusters: basic galactic building blocks throughout time and space', Proceedings of the International Astronomical Union, IAU Symposium, Volume 266, p. 293

\refer Vanbeveren, D., Belkus, H., Van Bever, J., Mennekens, N., 2009, Astrophys. Space Sci. 324, 271.

\refer Vanbeveren, D., De Donder, E., Van Bever, J., Van Rensbergen, W., De Loore, C., 1998, New Astronomy 3, 443.

\refer Vanbeveren, D., Herrero, A., Kunze, D., van Kerkwijk, M., 1993, Space Sci. Reviews 66, 395.

\refer Vanbeveren, D., Van Bever, J., Belkus, H., 2007, ApJ 662, 107.

\refer Vanbeveren, D., Van Rensbergen, W., De Loore, C., 1998b, in 'The Brightest Binaries', Kluwer Academic Pub., Dordrecht.

\refer von Zeipel, H., 1924, MNRAS 84, 665.

\refer Weidner, C., Kroupa, P., 2006, MNRAS 365, 1333.

\refer Yoon, S.C., Cantiello, M., 2010, ApJ 717, 62.

\refer Yungelson. L.R., van den Heuvel, E.P.J., Vink, J.S., et al., 2008, A\&A 477, 223.

\endrefer           
\end{document}